\documentclass[12pt]{article}
\usepackage{amsmath}
\usepackage{amssymb}
\usepackage{epsfig}
\usepackage{cite}
\usepackage{color,colordvi}

\begin{document}
\vspace{0.01cm}
\begin{center}
{\Large\bf  BICEP2 in Corpuscular Description of Inflation} 

\end{center}

\vspace{0.1cm}

\begin{center}

{\bf Gia Dvali}$^{a,b,c}$ and {\bf Cesar Gomez}$^{b,e}$\footnote{cesar.gomez@uam.es}

\vspace{.6truecm}


{\em $^a$Center for Cosmology and Particle Physics\\
Department of Physics, New York University\\
4 Washington Place, New York, NY 10003, USA}

{\em $^b$Arnold Sommerfeld Center for Theoretical Physics\\
Department f\"ur Physik, Ludwig-Maximilians-Universit\"at M\"unchen\\
Theresienstr.~37, 80333 M\"unchen, Germany}


{\em $^c$Max-Planck-Institut f\"ur Physik\\
F\"ohringer Ring 6, 80805 M\"unchen, Germany}


%

{\em $^e$
Instituto de F\'{\i}sica Te\'orica UAM-CSIC, C-XVI \\
Universidad Aut\'onoma de Madrid,
Cantoblanco, 28049 Madrid, Spain}\\

\end{center}


\begin{abstract}
\noindent  
 
{\small
A corpuscular quantum  description of inflation shows that there is no fundamental problem with 
trans-Planckian  excursions of the inflaton field up to about $100$ Planck masses, with the upper bound coming from the corpuscular quantum effects.  In this description the $r$-parameter measures  
the ratio of occupation numbers of gravitons versus inflatons, which, according to BICEP2, was roughly  a half at the time of  60 e-foldings prior to the end of inflation. We stress that in non-Wilsonian UV self-completion of gravity 
any trans-Planckian mode coupled to inflaton is a black hole.  Unlike the Wilsonian case, their integration-out gives an exponentially-suppressed effect and is unable to prevent  trans-Planckian excursions of the inflaton field.}

\end{abstract}

\thispagestyle{empty}
\clearpage

  The measurement of a large tensor mode by BICEP2 \cite{BICEP2}  gives a very strong indication that quantum gravity effects are reliably computable within the standard quantum field theory approach as long as the graviton wave-lengths are larger than the Planck length, $L_P$. 
   The standard computations for both tensor \cite{alexey} and scalar \cite{chibisov-mukhsnov} modes  are performed in the semi-classical limit, in which, although
the graviton fluctuations are quantized, the gravitational background itself is still treated classically.  
 In quantum language this means that the quantum corpuscular structure of the  background remains unresolved, or equivalently, the number of corpuscles $N$ is taken infinite.  This is very similar, for example, to interaction between an isolated quantum particle and a classical electromagnetic wave, in which the 
 number of constituent photons is taken infinite.   Obviously, in this treatment any $1/N$ effect coming from the finite occupation number of photons is left unaccounted. 
 Similarly, in the standard treatment of inflation any effect coming from the quantum sub-structure of the background remains unaccounted. 
    
    Thus, for a complete picture,  we must undertake one more step
  and formulate a corpuscular description \cite{usBH, us} which  resolves the would-be classical backgrounds, such as black holes or inflationary spaces,  in terms of constituent  quantum corpuscles, gravitons and inflatons.
 Such a picture is necessary for addressing  the fundamental questions that  
 BICEP2 results are posing.

 In particular, when interpreted within the inflationary paradigm \cite{Guth}, BICEP2 result 
   suggests that the range of change of the inflaton field is trans-Planckian, as it is the case for example in Linde's chaotic inflation \cite{linde} with the potential $V=m^2\phi^2$. 
  In particle physics one often raises the question about  the fundamental consistency of  trans-Planckian  changes of expectation values of  scalar fields.   For short, we shall refer to these expectation values 
  as VEVs, although the fields are not necessarily in the ground-state. 
   In order to answer this  question one needs a microscopic theory of inflation.  We shall address this issue within 
   a recently formulated corpuscular theory of inflation \cite{us}, and show that the trans-Planckian change of the inflaton VEV is permitted although the range is unexpectedly strongly limited from above.

 Within this description the large tensor mode translates as a relatively large ratio of the occupation numbers of  graviton and inflaton
 quanta that compose the would-be-classical background.   As long as the Hubble volume is  larger than the Planck length, the  de Broglie wave-lengths of constituent gravitons and inflatons are large and 
 they are weakly interacting. The standard results \cite{alexey}, \cite{chibisov-mukhsnov} then can be reliably reproduced from our corpuscular theory, as long as the following bound is satisfied \cite{us}
 \begin{equation} 
 \epsilon \, >  \, N^{-2/3} \, . 
 \label{boundeps} 
 \end{equation}
 Here $\epsilon$ is the standard inflationary slow-roll parameter and $N$ is the occupation number of soft gravitons that compose the classical background.  For  $m^2\phi^2$ potential this bound translates as, 
 \begin{equation}
 \phi \, < \, M_P (M_P/m)^{2/5} \, , 
 \label{bound} 
 \end{equation}
 where $M_P$ is the Planck mass. 
  This upper bound comes from corpuscular  effects \cite{us} and will be explained below. 
  From observations $m \sim  10^{13}$, and thus the upper bound is within the reliability range. 
  Notice, however, that  (\ref{bound}) is more stringent  than what we would get from imposing a naive semi-classical constraint on energy density to stay below the Planck density, which instead would give 
  $\phi \, < \, M_P (M_P/m)^{1/2}$.

  Let us now explain this point. For definiteness, following \cite{us}, we shall discuss  the simplest 
  large field inflationary model with potential $V = m^2\phi^2$ \cite{linde}.  It is well-known 
  that classically this system enters the inflationary regime whenever the slow-roll parameter 
  $\epsilon \, \equiv  \, M_P^2 /\phi^2$ is less than one. Or equivalently, if the inflaton VEV  exceeds the Planck 
  mass.  Is this a problem at the fundamental level?  
   We shall address this question within the corpuscular theory, which on one hand 
   reproduces all the known predictions, but at the same time is powerful enough to indicate 
   whether some inconsistency arises at the fundamental level.  
   
    Within this corpuscular description the above system, including the classical background, is viewed quantum-mechanically as being composed out of corpuscles, graviton and inflaton quanta. 
  That is, the would-be-classical scalar and gravitational fields are viewed as a multi-particle quantum state of 
  inflaton and graviton Bose-gasses. The  occupation numbers are given by 
  \begin{equation}
 N \, =\, {M_P^2 \over H^2} ~~~ {\rm and} ~~~ N_{\phi} = { N \over \sqrt{\epsilon}}
 \label{Ns}
 \end{equation}
  respectively.  Note that these corpuscles are in a bound-state and therefore are off-shell. 
   However, ``off-shellness" is mild and therefore it makes perfect 
 sense to talk about constituency.   One can draw an analogy\cite{usBH}  with the QCD baryon as composite of $N$-quarks with $N$ the number of colors, which admits a nearly-classical description\cite{Witten} for 
 large-$N$. 
  But,  unlike the baryon which is stable,  the Hubble patch is a {\it loose union}  and depletes quantum mechanically.  Unlike the quarks the gravitons and inflatons can exist in the free-state and 
 the Hubble patch depletes quantum mechanically by emitting them out of the would-be-classical background.    
 
  In this picture the classical evolution is recovered as a mean-field limit ($N=\infty$, $\hbar=0$, 
  $H=$ finite), whereas, the standard semi-classical physics of  \cite{alexey},\cite{chibisov-mukhsnov}
    up to $1/N$-corrections is recovered 
  for $N=$ finite, $\hbar=$ finite  and $H=$ finite.  
  
   Once we have a microscopic quantum theory we can fully handle the physical meaning 
    of the trans-Planckian inflaton VEV.   It is obvious from  (\ref{Ns}) that the physical meaning of $\phi \gg M_P$ is the relative increase of inflaton occupation number.   This increase would be problematic if it 
   could result into a breakdown of the predictive power by, for example, rendering some interactions strong
    or expansion parameters large.  
    
  In fact, the allowed range for $\phi$ VEV is big enough for what is needed for accommodating chaotic 
 inflation and deriving the density perturbations reliably  at the fundamental level.   On the other hand the 
 range is still strongly limited by (\ref{bound}). 
  
   In order to see this, let us focus on reproducing density perturbations. 
    In this framework the density perturbations are the result of quantum depletion. Namely, due to quantum re-scattering the constituent corpuscles are systematically pushed out of the would-be ground-state. 
     The hierarchy between the scalar and tensor modes of density perturbations is the result 
     of the different rates of scattering in graviton-inflaton and graviton-graviton channels.  This difference 
  is the simple consequence of the difference in occupation numbers. Since the background houses $1/\sqrt{\epsilon}$ times more inflatons than gravitons,  the depletion in the scattering channels with inflaton participation  is enhanced. Correspondingly,  we end up with the following expression for $r$-parameter,
 \begin{equation}
    r = {N^2 \over N_{\phi}^2} = \epsilon \,. 
 \label{r}
 \end{equation} 
 It is interesting that in the light of the BICEP2 data, suggesting large value  $r \simeq 0.2$, the above relation tells us that 
 at the time of 60 e-foldings prior the end of inflation the occupation numbers of inflatons and gravitons were  different only by a factor of two.  Whether there is a deeper physical meaning to this fact needs to be understood.

  On the other hand the computational range for $\phi$  is much more narrow than what one would naively expect from the semi-classical theory,  and is limited by the bound (\ref{boundeps}) \cite{us}, 
 which for $V=m^2\phi^2$ translates to (\ref{bound}).  
  
    This limit has no counter-part in the semi-classical analysis and is 
  intrinsically-quantum.  The reason are the corpuscular  corrections. As we said, the corpuscular picture 
  agrees with semi-classical treatment only up to $1/N$ corpuscular effects, per  Hubble time. 
   That is, the standard semi-classical treatment cannot fully account for the true 
   composite nature of the background and this amounts to an error of order $1/N$ per each Hubble time.
 These corrections keep accumulating over the duration of inflation.  The physical meaning of this growing  correction is very transparent.   Due to quantum depletion, after each  Hubble time the background is left with $\Delta N \sim 1/\sqrt{\epsilon}$ less number of corpuscles. This change is not accounted by the classical evolution.  Thus, for each subsequent Hubble time there is an error between the two computations
 (semi-classical and quantum).  
 
This corpuscular error sums up to order-one difference after the number of elapsed Hubble times becomes $\sim N\sqrt{\epsilon}$.   Should the inflation be allowed to last this long, the semi-classical treatment would 
simply become unreliable.  But, on 
 the other hand number of Hubble-times computed from the classical evolution is given by $\sim 1/\epsilon$. 
 Requiring the consistency between the two values creates a {\it reliability}  bound (\ref{boundeps}) or equivalently 
 (\ref{bound}).    Violation of this bound 
renders the standard semi-classical treatment fully unreliable, as it makes duration of inflation long enough 
for accumulation of order-one quantum error.  
 
   For the realistic $m^2\phi^2$-inflation, $N \sim 10^{12}$ and $1/N$ effects give a very small correction for 
   the duration of $60$ Hubble times. However, for the time-scales of order $\sim N\sqrt{\epsilon}$, the corpuscular corrections make a dramatic difference. 
   The upper limit  (\ref{boundeps}-\ref{bound}) on the inflaton VEV comes   precisely from such long 
   time-scale effects.  Although,  for a
  $60$-Hubble-times-lasting inflation this 
   bound has not much effect on standard predictions, it can dramatically change the large-scale 
   picture of the inflationary scenario and in particular exclude eternal inflation from the computational 
  range.  The detailed discussions in this direction can be found in \cite{us}. 
 
  Finally, let us make a very general point about the trans-Planckian excursions of the inflaton 
  field and the nature of UV-completion of gravity. 
  
   The general uneasiness with trans-Planckian excursions of the inflaton VEVs comes from the intuition that 
 during such excursions modes coupled to the inflaton (in non-derivative manner) 
 become trans-Planckian and their integration-out  
 can induce an effective potential that prevents such excursions. 
  We would like to stress, that such a worry is based on a conventional Wilsonian thinking and the story is 
  very different for non-Wilsonian self-completion \cite{self}.

  Although the corpuscular description of inflation \cite{us} and black holes \cite{usBH} does not {\it a priory} assume any concrete form of  UV-completion of gravity,  it is preparing a firm microscopic ground for the concept of 
  non-Wilsonian self-completion\cite{self}.  
  The key ingredient of self-completeness is to replace ultra-Planckian modes of energy $E$ by
  multi-particle states (black holes) of large occupation number $N(E)\sim E^2L_P^2$ of soft gravitons.
 
  In this picture, the isolated one-particle states of mass $M \gg M_P$ do not exist and they are replaced  
 by black holes, or equivalently, by the states composed of many soft gravitons.  Unlike the Wilsonian case,  integrating out such a multi-particle (almost-classical) state must result in an exponentially suppressed contribution to the 
 inflaton potential.    Such a contribution cannot prevent trans-Planckian excursions of the inflaton field. 
   We must stress that the analogous point was independently made by Kehagias and Riotto \cite{Kehagias} . 
  
  In conclusion,  non-Wilsonian self-completion of gravity suggests  a built-in defense mechanism for an approximate shift symmetry for large values of the scalar fields.

\section*{Acknowledgements}
 
The work of G.D. was supported by Humboldt Foundation under Alexander von Humboldt Professorship,  by European Commission  under 
the ERC Advanced Grant 226371 and ERC Advanced Grant 339169 ``Selfcompletion'',   by TRR 33 \textquotedblleft The Dark
Universe\textquotedblright\   and  by the NSF grant PHY-0758032. 
The work of C.G. was supported in part by Humboldt Foundation and by Grants: FPA 2009-07908, CPAN (CSD2007-00042) and HEPHACOS P-ESP00346 and by the ERC Advanced Grant 339169 ``Selfcompletion'' .

It is a pleasure to thank organizers of  the New York meeting under the project ``The Particle Physics and Cosmology of Supersymmetry  and String Theory", where some of our results were presented.

\end{document}